\documentclass[aps,pra,reprint,superscriptaddress]{revtex4-2}
\usepackage[colorlinks=true, allcolors=blue]{hyperref}
\usepackage{graphicx}
\usepackage{times}
\usepackage{bm}
\usepackage{amsmath,amssymb}
\usepackage{ifpdf}
\usepackage{multirow}

\begin{document}

\title{Applicability of the Dirac-Fock method combined with Core Polarization in calculations of alkali atoms}

\author{A.~A.~Bobylev}
\email[E-mail:]{artem.bobyljow@gmail.com}
\affiliation{Department of Physics, St. Petersburg State University,
Petrodvorets, Oulianovskaya 1, 198504, St. Petersburg, Russia}
\affiliation{Petersburg Nuclear Physics Institute named by B.P. Konstantinov of NRC ``Kurchatov Institute'', Orlova roscha 1, 188300 Gatchina, Leningrad region, Russia}

\author{J.~J.~Lopez-Rodriguez}
\affiliation{Department of Physics, St. Petersburg State University,
Petrodvorets, Oulianovskaya 1, 198504, St. Petersburg, Russia}

\author{P.~A.~Kvasov}
\affiliation{Department of Physics, St. Petersburg State University,
Petrodvorets, Oulianovskaya 1, 198504, St. Petersburg, Russia}

\author{M.~A.~Reiter}
\affiliation{Department of Physics, St. Petersburg State University,
Petrodvorets, Oulianovskaya 1, 198504, St. Petersburg, Russia}

\author{D.~A.~Solovyev}
\affiliation{Department of Physics, St. Petersburg State University,
Petrodvorets, Oulianovskaya 1, 198504, St. Petersburg, Russia}
\affiliation{Petersburg Nuclear Physics Institute named by B.P. Konstantinov of NRC ``Kurchatov Institute'', Orlova roscha 1, 188300 Gatchina, Leningrad region, Russia}

\author{T.~A.~Zalialiutdinov}
\affiliation{Department of Physics, St. Petersburg State University,
Petrodvorets, Oulianovskaya 1, 198504, St. Petersburg, Russia}
\affiliation{Petersburg Nuclear Physics Institute named by B.P. Konstantinov of NRC ``Kurchatov Institute'', Orlova roscha 1, 188300 Gatchina, Leningrad region, Russia}

\begin{abstract}
In this work, we investigate the applicability of the core-polarization-corrected Dirac–Fock method, formulated within the framework of the local Dirac–Hartree–Fock (LDF) potential, for the accurate determination of static scalar and tensor electric dipole polarizabilities. This work presents theoretical values of blackbody-radiation-induced Stark shifts of atomic energy levels. The Dirac–Fock method augmented by core-polarization corrections is employed not only to evaluate these shifts but also to compute the Bethe logarithm for alkali-metal atoms. The results are critically compared with data available in the contemporary literature, and the strengths and limitations of the present approach are discussed.
\end{abstract}
\maketitle
%
%
\section{Introduction}
%
Currently, calculations of atomic and molecular characteristics, such as polarizabilities, Stark shifts, hyperfine structure parameters, and QED corrections, are of great interest for various theoretical and experimental applications. The values of these quantities can be used in precision spectroscopy, in the development of atomic clocks, metrology, etc. \cite{Sahoo-AtCl,CI-calc,Kozlov-RevModPhys}.

Of particular interest are many-electron monovalent atoms and ions. These systems possess a deceptively simple electronic structure, characterized by a single valence electron outside a closed-shell core, yet they pose significant challenges for high-precision atomic-structure calculations. These difficulties arise primarily from electron correlation effects between the valence electron and the core electrons. Such correlations induce a redistribution of the core electron density, which significantly influences the magnitude of the polarizability and renders single-particle approaches—such as the Hartree–Fock method—inadequate for quantitative accuracy. In the case of heavy elements (e.g., Rb, Cs, Fr), an accurate treatment further requires the inclusion of relativistic corrections, as demonstrated in recent studies \cite{PhysRevA.77.012515,WU2024129805,PhysRevA.110.062805,PhysRevA.111.022816,CHEUNG2025109463}.

To achieve high-precision predictions of atomic properties—such as transition energies, static and dynamic polarizabilities, hyperfine structure splittings, and quantum electrodynamic (QED) radiative corrections—it is essential to employ rigorously defined {\it ab initio} theoretical frameworks. Such approaches must systematically incorporate electron correlation effects while avoiding reliance on empirical or adjustable parameters, thereby ensuring predictive power and transferability across atomic systems. Representative methods fulfilling these criteria include multiconfiguration self-consistent field (MCSCF) theory~\cite{andersson1992electric}, configuration interaction (CI) expansions of varying excitation rank~\cite{hibbert1975developments}, coupled-cluster (CC) techniques that account for single, double, and higher-order excitations~\cite{bishop1987coupled}, and multireference perturbation theories tailored for open-shell atoms~\cite{kozlov1999polarizabilities}. While these methods provide a robust foundation for quantitative atomic structure calculations, their implementation typically entails substantial computational cost and algorithmic complexity, particularly for heavy or highly correlated systems.

In addition to ab initio methods, there are semi-empirical approaches based on the use of model potentials. Semi-empirical methods effectively balance computational accuracy with the time required to calculate atomic characteristics. One such approaches is the "Dirac-Fock Core plus Polarization (DFCP)" method~\cite{yan:pr:2008,PhysRevA.94.032503}.
In this work, we present our variation of such an approach, based on the use of a local version of the Dirac-Hartree-Fock potential~\cite{PhysRevA.72.062105,PhysRevLett.94.213002}. Hereafter, we refer to such a method as LDFCP and investigate its applicability in calculating atomic characteristics. This article presents the results of calculations of polarizabilities, Stark energy shifts induced by blackbody radiation, and the quantity entering the electron self-energy correction (the Bethe logarithm). The motivation behind our research is that the LDFCP method is simple to implement and doesn't require extensive computational resources. Moreover, in the approximation of a one-electron state in an effective self-consistent field, it is possible to perform a direct summation over the intermediate states present in the aforementioned second order properties. Based on the results obtained, we provide a comparative analysis with the data available in modern literature and, consequently, discuss the validity of applying the LDFCP method.

The work is organized as follows. The next section~\ref{meth} provides a brief description of the applied method. Then in section~\ref{pol}, we present the formalism employed for the evaluation of static scalar and tensor electric dipole polarizabilities. These quantities are subsequently used in section~\ref{St} to compute blackbody-radiation-induced Stark shifts of atomic energy levels. Complementing this analysis, section~\ref{LogB} addresses the calculation of the Bethe logarithm, a key ingredient in the leading-order quantum electrodynamic correction to atomic binding energies, thereby providing a more complete characterization of radiative effects in the systems under study. Finally, the last part of the work provides the main conclusions and a comparative analysis of the LDFCP method in relation to similar results obtained using other approaches. Relativistic units ($\hbar=c=1$, $\hbar$ is the reduced Planck constant, and $c$ is the speed of light
) are used throughout the paper unless otherwise specified.

%
%

\section{Methods}
\label{meth}

Calculations in this work are carried out using an effective one-electron approximation. The valence electron is considered within a self-consistent field of a frozen core. The correlation interaction between the core electrons and the valence electron is taken into account using a semi-empirical core polarization potential. We employed the "Dirac-Fock plus Core Polarization"(DFCP) method~\cite{yan:pr:2008,PhysRevA.94.032503}, which is based on a local version of the Dirac-Hartree-Fock potential (LDF, see references~\cite{PhysRevA.72.062105,PhysRevLett.94.213002}). 

In this approach, the solution to the Dirac wave equation for the valence electron is found in standard form through the large, $g(r)$ and small $f(r)$ components:
\begin{eqnarray}
\label{1}
(-\mathrm{i}\boldsymbol{\alpha}\cdot\nabla + \beta m_e + V(\mathbf{r}))\psi_n(\mathbf{r}) = \varepsilon_n\psi_n(\mathbf{r}),
\\
\label{wfde}
\psi({\bf r})=
\frac{1}{r}\left(\begin{array}{c}
g(r)\;\Omega_{\kappa m}({\bf n})\\
\mathrm{i}f(r)\;\Omega_{-\kappa m}({\bf n})
\end{array}\right).
\end{eqnarray}
Here, $\Omega_{\kappa m}$ represents spherical tensors that depend on the angular quantum number $\kappa$ and the projection of the total angular momentum $m$. Eigenvalues are denoted by $\varepsilon_n$, $m_e$ is the electron mass, and $\boldsymbol{\alpha}$, $\beta$ are the Dirac matrices. Vectors are given in bold with the notations ${\bf n} = {\bf r}/r$ and $\nabla$ for the gradient operator.

The radial part of the Dirac equation, Eq.~(\ref{1}), and its solution are
\begin{eqnarray}  \label{radp}
H_{\kappa}\phi=E\phi,
\end{eqnarray}
where
\begin{eqnarray}\label{hamilt}
H_{\kappa}=
\begin{pmatrix}
m_e+V(r) & -\frac{d}{dr}+\frac{\kappa}{r} \\
\frac{d}{dr}+\frac{\kappa}{r} & -m_e+V(r)
\end{pmatrix},
\\
\label{slcomp}
\phi({ r})=
\left(\begin{array}{c}
g(r)\\
f(r)
\end{array}\right).
\end{eqnarray}
The potential involved in the eigenvalue problem, $V(r)$, is equal to 
\begin{eqnarray}\label{pot}
V(r) = V_{\rm LDF}(r) +V_{\rm CP}(r),
\end{eqnarray}
where $V_{\rm LDF}(r)$ is the local Dirac-Hartree-Fock potential, which sets the self-consistent field of the frozen core. 

A semi-empirical potential $V_{\rm CP}(r)$ describes the valence electron interaction with core. It can be defined, see, for example, \cite{yan:pr:2008,PhysRevA.94.032503}, as
\begin{eqnarray} \label{cp}
V_{\rm CP}(r) = -\frac{\alpha_c}{2\,r^4}\,\bigl(1-e^{-r^6/\rho_{\kappa}^6}\bigr)\,,
\end{eqnarray}
where $\alpha_c$ is the static electronic polarizability of the core, and $\rho_{\kappa}$ is a cutoff parameter that allows the potential to be finite at the origin. The parameter $\rho_{\kappa}$ depends on the quantum numbers $\kappa=(j+1/2)(-1)^{l+j+1/2}$ where  $l$ is the electron orbital momentum and $j$ is the total angular momentum of the valence electron. It can be chosen in such a way that the eigenvalues obtained by solving equation (\ref{radp}) correspond to the experimental values of the atom's energies ($ns$, $np_{j}$, $nd_{j}$ $\dots$), taken from the NIST database. 

The potential $V_{\rm CP}$ introduces a correction to the dipole electrical transition operator, as a result of which it is modified to 
\begin{eqnarray} \label{troper}
 r \longrightarrow r\left(1 -\frac{\alpha_c}{\,r^3}\,\left(1-e^{-r^6/\rho_{}^6}\right)^{1/2}\right).
\end{eqnarray}
Here $\rho=(\rho_{\kappa_{i}}+\rho_{\kappa_{f}})/2$, $i$ and $f$ correspond to the initial and final atomic states, respectively. The core static polarizability $\alpha_c$ values can be taken from literature \cite{article} and are collected in Table~\ref{tab:1} for the alkali elements.
\begin{table}[ht!]
\caption[]{The core static polarizability. The values are given in atomic units (a.u.), see \cite{article}.} 
\label{tab:1}
\resizebox{0.8\columnwidth}{!}{
\begin{tabular}{l | l | l | l | l | l}
\hline
\hline
Li & Na & K & Rb & Cs & Fr \\

\hline
$0.192486$ & $0.9947$ & $5.354$ 
& $9.1\,\,\,\,$& $15.81$ & $20.4$ \\
\hline
\hline
\end{tabular}
}
\end{table}

To solve the Dirac equation (\ref{radp}), the large $g(r)$ and small $f(r)$ components of the wave function are expanded over a finite basis set constructed using B-splines \cite{PhysRevA.37.307}, while imposing the dual-kinetic-balance (DKB) conditions \cite{PhysRevLett.93.130405}. Application of the DKB approach is implied for the extended charge nucleus only. The point-like nucleus case can be accessed by the extrapolation of the extended-nucleus results in vanishing nuclear size. The main advantage of the DKB method is the reproduction of the complete pseudo-spectrum of electron energies, consisting of a finite discrete set. This allows reducing the problem to calculating finite sums, which are then computed numerically (the so-called sum-over-states method). The pseudo-spectrum is constructed so that the lower few positive Dirac energy states precisely correspond to the actual low-lying bound states.

Therefore, the direct procedure – summing over bound states and integrating over continuum states within the perturbation theory – is replaced by finite sums over the pseudo-spectrum. This is particularly important in our work, as the calculated quantities involve sums over intermediate states.
All calculations were performed using systematically increased box size and basis sets (130 - 230 basis function). The box size was chosen sufficiently large to ensure convergence of the calculated values, including values for which highly excited states must be taken into account. 
We managed to reproduce a pseudo-spectrum containing approximately $30 - 40$ bound states.

\section{
Scalar and tensor polarizabilities}
\label{pol}

The interaction of an atom with a classical external electric field $\bf{E}$ gives rise to shifts in its energy levels, known as the Stark effect. This shift can be estimated within the framework of second-order perturbation theory. We are interested in the quadratic effect, as subsequent calculations pertain to the case of isotropic external thermal radiation \cite{PhysRevLett.42.835,PhysRevA.23.2397}.The perturbation caused by an external field can be  described by a multipole expansion of the perturbing potential. For the $2^\ell$-pole order, see \cite{patil2012asymptotic}, assuming that the electric field strength vector $\bf{E}$ is directed along the $z$ axis, the perturbation operator takes the form
\begin{eqnarray}
 \delta V_{\ell} = -E \sum_i r_i^{\ell} {C}^{(\ell)}_{0}(\mathbf{\hat{r}}_i)
\label{deltaV}
\end{eqnarray}
where ${C}^{(\ell)}_m(\mathbf{\hat{r}})$ is a normalized spherical harmonics. 

The second‑order energy shift induced by this perturbation is expressed through the static $2^\ell$-pole polarizability $a_{\ell}$:
\begin{eqnarray}
 \delta E^{(2)}_{\ell} = -\frac{1}{2} \, \alpha_{\ell} \, E^2
\label{deltaE}
\end{eqnarray}
The static $2^\ell$-pole polarizability comprises three  contributions
\begin{eqnarray}
\alpha^{(\ell)}_{0} = \alpha^{(\ell)}_{\rm v} + \alpha^{(\ell)}_{\rm c} + \alpha^{(\ell)}_{\rm cv},
\label{alpha_decomp}
\end{eqnarray}
where $\alpha^{(\ell)}_{\rm v}$ is the valence part, $\alpha^{(\ell)}_{\rm c}$ arises from the core electrons, and $\alpha^{(\ell)}_{\rm cv}$ - the valence-core coupling term.
The valence contribution of $\alpha^{(\ell)}_{\rm v}$ is calculated using the formula:
\begin{eqnarray}
\alpha^{(\ell)}_{\rm v} = \frac{2}{(2\ell+1)(2j_a+1)}
\sum_{n } \frac{ |\langle a \|  r^{\ell} {C}^{(\ell)}_{0}(\mathbf{\hat{r}}) \| n \rangle|^2 }{{E_{n}-E_a}},
\label{alpha_v}
\end{eqnarray}
Scalar dipole polarizability ($\ell=1$) calculations were performed. The valence part was calculated using the formula (\ref{alpha_v}), the core part, which is taken from Table~\ref{tab:1},  and the valence-core coupling term, which is neglected because it is too small.

For atomic states with total angular momentum $J > 1/2$, the polarizability possesses not only a scalar  part but also a tensor component that governs the relative splitting of magnetic substates with different values of the magnetic quantum number $M$. The tensor part of the dipole ($\ell=1$) polarizability is given by
\begin{eqnarray}
\label{TPol}
\alpha_{2}=4\left(\frac{5j_a(2j_a-1)}{6(j_a+1)(2j_a+1)(2j_a+3)}\right)^{1/2}
\\
\nonumber
\times\sum\limits_n
  (-1)^{j_a+j_n}
  \left\{ \begin{array}{ccc}
   j_a & 1 & j_n \\
   1 & j_a & 2
   \end{array} \right\}
  \frac{ |\langle a \| {\bf r}\| n  \rangle|^2 } {E_{n}-E_a}.
\end{eqnarray}

The summation in Eqs.~(\ref{alpha_v}), (\ref{TPol}) is carried out over the entire spectrum of intermediate states of the atomic system (including integration over continuum states). According to the DKB method used, the sum over $n$ is replaced by a summation over the pseudo-spectrum of energy states obtained for solving the Dirac equation in a box of finite size.
Expressions (\ref{alpha_v}) and (\ref{TPol}) contain the reduced matrix elements of the dipole moment operator, which are derived in accordance with the Wigner-Eckart theorem \cite{varshalovich1988quantum}.

The values of the atomic scalar polarizabilities obtained in the present work and those taken from the literature are listed in Tables~\ref{tab:2}-\ref{tab:7}. The latter represent calculations based on the configuration interaction method and are extendable to utilize second-order perturbation theory or the all-order method to attain even higher accuracy \cite{2025pCI,KIRUGA2026109951}. The numerical results for the scalar polarizability of neutral alkali atoms are given in Tables~\ref{tab:2} and \ref{tab:3}. Table~\ref{tab:5} reproduces the results of scalar polarizability calculations for highly excited (Rydberg) states, which are compared with analogues from the work \cite{PhysRevA.94.032503}, obtained by the DFCP method. Tables~\ref{tab:6} and \ref{tab:7} provide the results of calculations for tensor polarizability.

By comparing our results with published data, we can conclude that the polarizabilities obtained using the LDFCP method are consistent with high-precision calculations, differing by one percent or less. It should be noted that the calculations within the framework of our approach do not require high computational costs and relate to the unambiguous accounting of the entire complete Dirac spectrum when summing over intermediate states. The main reason for the deviations of the results from those established in modern literature is the semi-empirical potential. In our opinion, the obtained accuracy is sufficient given the current level of experimental precision.

\begin{table}[ht!]
	\caption[]{Static electric dipole scalar polarizabilities, $\alpha_0$, for neutral alkali atoms (designated at the top of each cell), are listed in atomic units. The each first row presents values obtained using the LDFCP method, the second row (for Li and Cs atoms) shows data without the CP correction, while the lower row provides values from the work of \cite{KIRUGA2026109951}. The presented numerical values are grouped according to the principal quantum number for various $n^{2s+1}S_J$, $n^{2s+1}P_J$, and $n^{2s+1}D_J$ states.}
    \label{tab:2}
	\begin{center}
		\begin{ruledtabular}
			\begin{tabular}{llllll}
				$n$ &   \multicolumn{1}{c}{$n^2S$}      &   \multicolumn{1}{c}{$n^2P_{1/2}$}   &     \multicolumn{1}{c}{$n^2P_{3/2}$}
				&     \multicolumn{1}{c}{$n^2D_{3/2}$} &     \multicolumn{1}{c}{$n^2D_{5/2}$}\\
				\hline\\[-0.2cm]
\multicolumn{6}{c}{\bf Li} \\[0.2cm]%
				2 &  163.8 & 126.8 & 126.8 &       &      \\
                 &  164.9 & 126.6 & 126.7 &       &      \\
				  & 164    & 126.89 & 126.91   &        &          \\[0.1cm]
				3 &   4128.4 & 27467 & 27471 & -14456 & -14458 \\
                 &  4131.7 & 27467 & 27471 & -14457 &-14458     \\
			       & 4130.2 &  28239   & 28237  & -14916  & -14914   \\[0.1cm]
				4 &  35336 & 262988 & 263020 & 4374090 & 4386700 \\
                 & 35343 & 262987 &263019 & 4374090 & 4386650     \\
				   & 35281  & 273260   &  273230  & 4923700  & 4936000       \\[0.1cm]

\hline\\[-0.2cm]
\multicolumn{6}{c}{\bf Na}\\[0.2cm]

				3 &  161.7 & 360.5 & 362.2 & 6438.4 & 6413.9 \\
				   & 162.44  & 359.94 & 361.63 & 6419.7 & 6395.3  \\[0.1cm]
				4 &  3079.7 & -4520.1 & -4450.0 & 634577 & 633699 \\
				   & 3100.3  & -4493.6 & -4423.5 & 635610 & 634840      \\[0.1cm]
				5 & 21539 & -58981 & -58387 & 4042970 & 4037220 \\
			  	 & 21697  & -58620  & -58027 & 4061600 & 4056700     \\[0.1cm]
\hline\\[-0.2cm]
\multicolumn{6}{c}{\bf K}\\[0.2cm]
                3 &      &        &        & 1429 & 1415 \\
				  &      &       &       & 1422      & 1408      \\[0.1cm]
				4 &  286.9 & 607.8 & 618.5 & 35897 & 35703 \\
				  & 290.25 & 602.1  & 612.7 & 35990 & 35810       \\[0.1cm]
				5 &  4906 & 7027 & 7207 & 193843 & 192748 \\
				   &  4961      &  7023      &  7202       & 195000      &  194000    \\[0.1cm]
				6 &  32470 & 43088 & 44326 & 712560 & 708266 \\
			  	 & 32840  & 43160     & 44390       &  718000     & 714000      \\[0.1cm]
			\end{tabular}
		\end{ruledtabular}
	\end{center}
\end{table}
\begin{table}[ht!]
	\caption{Table~\ref{tab:2} continued.}
     \label{tab:3}
	\begin{center}
		\begin{ruledtabular}
			\begin{tabular}{llllll}
				$n$ &   \multicolumn{1}{c}{$n^2S$}      &   \multicolumn{1}{c}{$n^2P_{1/2}$}   &     \multicolumn{1}{c}{$n^2P_{3/2}$}
				&     \multicolumn{1}{c}{$n^2D_{3/2}$} &     \multicolumn{1}{c}{$n^2D_{5/2}$}\\
				\hline\\[-0.2cm]
\multicolumn{6}{c}{\bf Rb} \\[0.2cm]
                4 &       &        &         & 586 &  553 \\
				 &         &         &        & 587      & 553         \\[0.1cm]
				5 &  314.7 &  807.7 &  869.2 & 17806 &  17422 \\
				 &  318.28       & 805        & 868      &18120   &  17740       \\[0.1cm]
				6 &   5127 &  12365 & 13361 & 92700 & 90345 \\
			     & 5167     & 12270        & 13270        &  94800      & 92400      \\[0.1cm]
				7 &   32408 &   83363 &  90172 & 323139 & 313215 \\
				 & 32620        & 83150       & 90230      & 332900        & 322900       \\[0.1cm]

\hline\\[-0.2cm]
\multicolumn{6}{c}{\bf Cs}\\[0.2cm]
                5 &       &        &       & -327 &  -423 \\
                &       &        &       & -386 &  -488 \\
				  &         &        &         & -328       & -433       \\[0.1cm]
				6 &  396.7 & 1307 &  1606 & -7030 & -9837 \\
                & 455.7 & 1399 &  1716 & -7200 & -10034\\
				  & 399.57       &  1343      & 1655        &   -5620  &  -83300     \\[0.1cm]
				7 &  6162 &  31308 &  39256 & -84121 & -108868 \\
                  &  6290 &  31629 &  39647 & -84458 & -109255 \\
				   &  6232      & 29880       & 37510   &-66600  &  -88800     \\[0.1cm]
				8 &  37681 &  244051 &  312101 & -469053 & -593199 \\
                &  37902 &  244703 &  312896 & -469623 & -593814 \\
			     & 38270      &  223300      & 284500        & -368900       & -475500      \\[0.1cm]

\hline\\[-0.2cm]
\multicolumn{6}{c}{\bf Fr}\\[0.2cm]
                6 &         &         &        & -249 & -584\\
				  &     &      &        & -259  & -613  \\[0.1cm]
				7 &  316.1 & 1174 &  2178 & -1979 & -9224 \\
				  & 316.3 &  1177       &  2213      &  -250     &  -7420     \\[0.1cm]
				8 &  4644 &  24118 & 45740 & -36090 & -91054 \\
				   & 4765      &  22670      & 43370       & -19400      & -73200      \\[0.1cm]
				9 &  25630 & 173590 &  338621 & -224560 & -478679 \\
			   & 26400       &  160770    & 315200       &  -138800     & -385600     \\[0.1cm]
				\end{tabular}
		\end{ruledtabular}
	\end{center}
\end{table}


\begin{table}[ht!]
	\caption{Tensor electric dipole polarizabilities, $\alpha_2$, in atomic units. For each principal quantum number $n$, the first and second rows correspond to values obtained using the LDFCP method and from \cite{KIRUGA2026109951}, respectively. Notations are consistent with Table~\ref{tab:2}.}
    \label{tab:6}
	\begin{center}
		\begin{ruledtabular}
			\begin{tabular}{llll}
				$n$ & \multicolumn{1}{c}{$n^2P_{3/2}$}
				& \multicolumn{1}{c}{$n^2D_{3/2}$} & \multicolumn{1}{c}{$n^2D_{5/2}$}\\
				\hline\\[-0.2cm]
\multicolumn{4}{c}{\bf Li} \\[0.2cm]
				2 & 1.53 & & \\
			       & 1.61 & & \\[0.1cm]
				3 & -2088 & 11078 & 15829 \\
			       & -2167 & 11400 & 16283 \\[0.1cm]
				4 & -19644 & -794916 & -1138700 \\
				   & -20708 & -901800 & -1291000 \\[0.1cm]

\hline\\[-0.2cm]
\multicolumn{4}{c}{\bf Na}\\[0.2cm]
                3 & -89.25 & -3587.5 & -5090.1 \\
				   & -88.38 & -3574.6 & -5071.8 \\[0.1cm]
				4 & -174.9 & -149169 & -212575 \\
				   & -167.7 & -149260 & -212740 \\[0.1cm]
				5 & 2454 & -925184 & -1318620 \\
			  	   & 2488 & -928210 & -1323300 \\[0.1cm]
\hline\\[-0.2cm]
\multicolumn{4}{c}{\bf K}\\[0.2cm]
                3 & & -487 & -678 \\
				   & & -484 & -674 \\[0.1cm]
				4 & -111.8 & -7776 & -10908 \\
				   & -108.7 & -7810 & -10960 \\[0.1cm]
				5 & -1087 & -37899 & -53001 \\
				   & -1063 & -38200 & -53400 \\[0.1cm]
			\end{tabular}
		\end{ruledtabular}
	\end{center}
\end{table}

\begin{table}[ht!]
	\caption{Table~\ref{tab:6} continued.}
    \label{tab:7}
	\begin{center}
		\begin{ruledtabular}
			\begin{tabular}{llll}
				$n$ & \multicolumn{1}{c}{$n^2P_{3/2}$}
				& \multicolumn{1}{c}{$n^2D_{3/2}$} & \multicolumn{1}{c}{$n^2D_{5/2}$}\\
				\hline\\[-0.2cm]
\multicolumn{4}{c}{\bf Rb} \\[0.2cm]
				4 & & -58 & -35 \\
				  & & -59 & -35 \\[0.1cm]
				5 & -169.1 & -1206 & -1087 \\
				  & -166.6 & -1310 & -1240 \\[0.1cm]
				6 & -2079 & 28 & 4168 \\
				  & -2060 & -530 & 3490 \\[0.1cm]

\hline\\[-0.2cm]
\multicolumn{4}{c}{\bf Cs}\\[0.2cm]
                5 & & 346 & 650 \\
				  & & 356 & 675 \\[0.1cm]
				6 & -260 & 9365 & 18426 \\
				  & -262 & 8736 & 17290 \\[0.1cm]
				7 & -4588 & 79421 & 158795 \\
				  & -4409 & 71040 & 141600 \\[0.1cm]
\hline\\[-0.2cm]
\multicolumn{4}{c}{\bf Fr}\\[0.2cm]
                6 & & 235 & 863 \\
				  & & 233 & 890 \\[0.1cm]
				7 & -454.2 & 4547 & 19883 \\
				  & -453.9 & 4032 & 18480 \\[0.1cm]
				8 & -6776 & 35254 & 157224 \\
				  & -6599 & 31120 & 143100 \\[0.1cm]
			\end{tabular}
		\end{ruledtabular}
	\end{center}
\end{table}


\begin{table}[ht!]
	\caption{Static electric dipole scalar polarizabilities in a.u. for Rydberg states. Each first row contains values obtained by the LDFCP method, the second row contains values from the work \cite{PhysRevA.94.032503}. The number in square brackets indicates the power of ten, $\times 10^{[\cdot]}$.}
     \label{tab:5}
	\begin{center}
		\begin{ruledtabular}
			\begin{tabular}{llllll}
				$n$ &   \multicolumn{1}{c}{$n^2S$}      &   \multicolumn{1}{c}{$n^2P_{1/2}$}   &     \multicolumn{1}{c}{$n^2P_{3/2}$}
				&     \multicolumn{1}{c}{$n^2D_{3/2}$} &     \multicolumn{1}{c}{$n^2D_{5/2}$}\\
				\hline\\[-0.2cm]
\multicolumn{6}{c}{\bf Rb} \\[0.2cm]
				8 &   0.132$\,$[6]       &  0.361$\,$[6]       &  0.392$\,$[6]       & 0.899$\,$[6]       &  0.867$\,$[6]        \\
				  &    0.132$\,$[6]       &  0.361$\,$[6]       &  0.392$\,$[6]       & 0.899$\,$[6]       &  0.867$\,$[6]              \\[0.1cm]
				10 &   0.109$\,$[7]      &  0.329$\,$[7]       &  0.358$\,$[7]       & 0.463$\,$[7]       &  0.442$\,$[7]       \\
			       &  0.109$\,$[7]      &  0.329$\,$[7]       &  0.358$\,$[7]       & 0.463$\,$[7]       &  0.442$\,$[7]       \\[0.1cm]
				12 &   0.523$\,$[7]      &   0.173$\,$[8]      &  0.188$\,$[8]       & 0.169$\,$[8]       &  0.159$\,$[8]       \\
				   &      0.523$\,$[7]      &   0.172$\,$[8]      &  0.188$\,$[8]       & 0.196$\,$[8]       &  0.159$\,$[8]       \\[0.1cm]
				15 &   0.317$\,$[8]      &  0.116$\,$[9]       &  0.127$\,$[9]      &  0.796$\,$[8]       &  0.740$\,$[8]       \\
			   	   &    0.317$\,$[8]      &  0.116$\,$[9]       &  0.127$\,$[9]      &  0.796$\,$[8]       &  0.740$\,$[8]        \\[0.1cm]
				20 &   0.285$\,$[9]      &  0.120$\,$[10]      &  0.131$\,$[10]     & 0.567$\,$[9]       &  0.516$\,$[9]           \\
				   &   0.289$\,$[9]      &  0.118$\,$[10]      &  0.130$\,$[10]     & 0.631$\,$[9]       &  0.582$\,$[9]       \\[0.1cm]
				30 &   0.554$\,$[10]     &  0.275$\,$[11]      &  0.303$\,$[11]      & 0.087$\,$[11]     &  0.772$\,$[10]    \\
				   &   0.554$\,$[10]     &  0.269$\,$[11]      &  0.297$\,$[11]      & 0.101$\,$[11]     &  0.905$\,$[10]      \\[0.1cm]
\hline\\[-0.2cm]
\multicolumn{6}{c}{\bf Cs}\\[0.2cm]
				9 &  0.151$\,$[6]       &  0.114$\,$[7]       &  0.148$\,$[7]       & $-$0.179$\,$[7]       & $-$0.224$\,$[7]       \\
				  &   0.154$\,$[6]       &  0.105$\,$[7]       &  0.136$\,$[7]       & $-$0.146$\,$[7]       & $-$0.185$\,$[7]    \\[0.1cm]
				10 &  0.466$\,$[6]       &  0.398$\,$[7]       &  0.520$\,$[7]       & $-$0.543$\,$[7]       & $-$0.674$\,$[7]      \\
				   &  0.477$\,$[6]       &  0.356$\,$[7]       &  0.463$\,$[7]       & $-$0.436$\,$[7]       & $-$0.548$\,$[7]     \\[0.1cm]
				12 &  0.278$\,$[7]       &  0.281$\,$[8]       &  0.371$\,$[8]       & $-$0.320$\,$[8]       & $-$0.395$\,$[8]      \\
			  	   &  0.286$\,$[7]       &  0.246$\,$[8]       &  0.321$\,$[8]       & $-$0.254$\,$[8]       & $-$0.315$\,$[8]   \\[0.1cm]
				15 &  0.203$\,$[8]       &  0.245$\,$[9]       &  0.325$\,$[9]       & $-$0.239$\,$[9]       & $-$0.293$\,$[9]      \\
				   &   0.211$\,$[8]       &  0.210$\,$[9]       &  0.275$\,$[9]       & $-$0.188$\,$[9]       & $-$0.231$\,$[9]       \\[0.1cm]
				20 &  0.214$\,$[9]       &  0.316$\,$[10]      &  0.421$\,$[10]      & $-$0.270$\,$[10]      & $-$0.329$\,$[10]     \\
				   & 0.226$\,$[9]       &  0.267$\,$[10]      &  0.350$\,$[10]      & $-$0.212$\,$[10]      & $-$0.258$\,$[10]   \\[0.1cm]
				30 &   0.476$\,$[10]      &  0.883$\,$[11]      &  1.18$\,$[11]      & $-$0.671$\,$[11]      & $-$0.813$\,$[11]    \\
				   & 0.511$\,$[10]      &  0.740$\,$[11]      &  0.972$\,$[11]      & $-$0.526$\,$[11]      & $-$0.636$\,$[11]    \\
			\end{tabular}
		\end{ruledtabular}
	\end{center}
\end{table}

The tables also include calculations performed without the core polarization correction for some atoms. This allows us to analyze the role of the semi-empirical correction in the Local-Dirac-Fock potential. In particular, the scalar polarizability was calculated for Li and Cs atoms. For the light atom, see Table~\ref{tab:2}, the core polarization correction is largely insignificant, not exceeding the stated accuracy of the method, i.e. $\lesssim 1\%$. In contrast, for the Cs atom (see, for example, the $nD$ states in Table~\ref{tab:3}) this correction has a noticeable impact on the final polarizability value.

Given the metrological importance of cesium and its use in atomic clocks, the following should be noted. Our calculated static polarizability for the $6^2S$ state of Cs shows much better agreement with the results of \cite{KIRUGA2026109951} than with \cite{chakraborty2023high}. However, the value $\alpha_0 = 401.04$ a.u. for the $6^2S$ state from \cite{chakraborty2023high} is itself in good agreement with both our result and that of \cite{KIRUGA2026109951}, with deviations not exceeding $1\%$. For the cesium $7^2S$, $7^2P_{1/2}$, and $7^2P_{3/2}$ states, our results for the static polarizability agree reasonably well with the theoretical work of \cite{KIRUGA2026109951}. The discrepancy is more significant, however, when compared to experimental measurements $6207.9$ \cite{PhysRevLett.132.233201}, $29660$ \cite{PhysRevA.89.042512} and $37280$ \cite{PhysRevA.89.042512}, respectively.

Regarding the tensor polarizability (see Tables~\ref{tab:6}, \ref{tab:7}), the LDFCP method exhibits reasonable convergence with the data from \cite{KIRUGA2026109951}. The most significant discrepancies emerge for the $nD$ states in all alkali elements studied and amount to approximately $10\%$. It should be noted, however, that calculations of the tensor polarizability within alternative methods carry significant uncertainty, leading to a possible spread in numerical results on the order of $100\%$ \cite{PhysRevA.83.052508}. Thus, our results (for instance, for the $6D_{3/2}$ state in rubidium) generally fall within this numerical uncertainty range. Moreover, given that the numerical error of our calculation is less than $1\%$, as confirmed by an analysis of other values, the obtained numbers can be expected to serve as reliable benchmark values.

Observing deviations in the values of the dipole tensor polarizabilities, we performed calculations of these quantities for Rydberg states of Cs and Rb atoms. A comparison of the obtained results with those from \cite{PhysRevA.94.032503} is presented in Table~\ref{tab:5}. In particular, it clearly shows good overall agreement, although significant deviations exist for the $30^2D_{3/2(5/2)}$ states in both atoms. A similar, albeit weaker, discrepancy is already observed for the $20^2D_{3/2(5/2)}$ states, attributed to the selection of CP correction parameters. The latter were chosen to achieve the best agreement between the calculated (within the employed method) and the NIST data energy values for the low-lying states.

We also note that the calculation method presented in section~\ref{meth} enables the study of the atomic polarizability's dependence on the finite nuclear size correction. Having performed corresponding calculations for different nuclear models (Fermi and shell models), we found no significant effect from this correction. Its magnitude is considerably smaller than the stated accuracy of the method; as expected for Rydberg states, it alters only the last digit reported in the tables.

%
%

\section{Stark shift of atomic level induced by blackbody radiation}
\label{St}
Currently, it is a well-known fact that the effect of the blackbody radiation (BBR) field on an atom leads to a shift of its energy levels. Due to the isotropy of thermal radiation, the energy shift is determined in the second order of time-dependent perturbation theory \cite{sobel2016introduction}. Given that the electric field strength, see, e.g., \cite{PhysRevA.23.2397}, of black body radiation is determined through the BBR intensity (in r.u.)
$\langle {E}^2(\omega)\rangle = (8/\pi)\omega^3(e^{\beta \omega}-1)^{-1}$, it can be found that
\begin{eqnarray}
\label{dStark}
\begin{aligned}
 \Delta E_a^{\mathrm{BBR}}=\frac{2}{3 \pi\left(2 j_a+1\right)} \sum_{n\neq a} \text { P.V. } \int_0^{\infty} d \omega\, n_\beta(\omega) \omega^3 \\
 \quad \times|\langle a| \boldsymbol{d}| n\rangle\left.\right|^ 2\left[\frac{1}{E_n-E_a+\omega}+\frac{1}{E_n-E_a-\omega}\right].
 \end{aligned}
\end{eqnarray}
In the formulas above $n_\beta(\omega)=1/(\exp(\beta \omega)-1)$ is the Planck distribution, $\beta=1/k_{\rm B}T$ ($k_{\rm B}$ is the Boltzmann constant, $T$ is the temperature in Kelvin), $\omega$, and P.V. denotes integration over frequency in the sense of the Cauchy principal value.

\begin{figure}[ht!] 
\centering
\includegraphics[width=0.5\columnwidth]{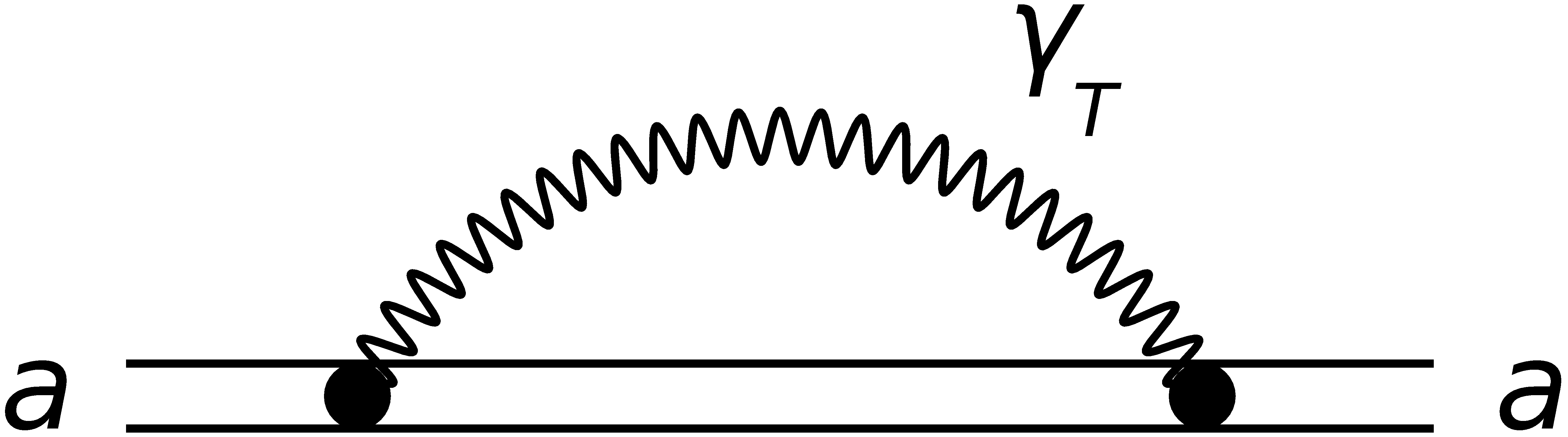} 
\caption{Thermal one-loop correction to the self-energy of an atomic electron at an arbitrary level $a$. Double line means the bound electron states in the Furry picture, the wavy line refers to the photon propagator. The notation $\gamma_T$ indicates that the thermal photon propagator is considered \cite{DHR,Don}.}
\label{fig:2}
\end{figure}

Derived within the one-loop QED approach for bound states at finite temperature \cite{PRA_2015,solovyev2020thermal}, this expression follows from considering the self-energy diagram of a bound electron, see Fig.~\ref{fig:2}.
The essential modification is the replacement of the ordinary photon propagator with a thermal propagator \cite{DHR,Don}. Then, it is straightforward to show (for example, based on the Sokhotsky-Plemel theorem) that the correction to the energy level $a$, obtained from this diagram, consists of real and imaginary parts. The real part in the leading order of multipole expansion exactly corresponds to the dynamic Stark shift Eq.~(\ref{dStark}), while the imaginary part determines the width of the level induced by blackbody radiation \cite{PRA_2015,solovyev2020thermal,vzqf-8261}.

The approach validated in the preceding section allows for the direct integration over frequency in expression (\ref{dStark}) describing the thermally induced dynamic Stark effect. The principal importance of this integration should be noted. It is underscored by recent fully relativistic calculations of thermally induced ac-Stark shifts for hydrogen and hydrogen-like ions presented in \cite{reiter2025}. Deviations of over $1\%$ from the standard quantum-mechanical values \cite{PhysRevA.23.2397} were observed when using a 'direct' numerical integration scheme. This discrepancy is primarily attributable to the unaccounted fine structure of atomic levels in \cite{PhysRevA.23.2397}, but is also contributed by the approximations made in that early work for evaluating the principal value integral.

Another well-established calculation method involves expanding the dynamic atomic polarizability in a series for small frequencies. These frequencies are determined by the temperature entering the Planck distribution. The calculations then reduce to evaluating the static polarizability and the corresponding dynamic corrections. Conversely, one can use a Taylor series expansion in terms of the small energy difference. However, these approximations fail when the frequency at the maximum of the blackbody distribution is comparable to the energy differences in the denominators of Eq.~(\ref{dStark}). Thus, direct numerical integration of the frequency integral appears to be the most suitable approach for calculating the thermal ac-Stark effect.

We performed calculations of the Stark shift at a temperature of $300$ Kelvin. The results of our computations for various states of elements in the alkali metal group of the periodic table are presented in Tables~\ref{tab:10}, \ref{tab:11}.
\begin{table}[ht!]
	\caption{Thermal Stark shifts at a temperature of $300$ K, values given in Hz. Each upper row contains values obtained by the LDFCP method, the second row contains values borrowed from the work \cite{PhysRevA.23.2397}. Standard notation is used for atomic states: $n$ is the principal quantum number, the superscript denotes multiplicity, determined by the spin moment, $S, P, D$ are the corresponding orbital momenta, and the subscript represents the total angular momentum. For Li and Cs, every second row shows the shift values obtained without the CP correction.}
    \label{tab:10}
	\begin{center}
		\begin{ruledtabular}
			\begin{tabular}{llllll}
				$n$ &   \multicolumn{1}{c}{$n {}^2S_{1/2}$}      &   \multicolumn{1}{c}{$n {}^2P_{1/2}$}   &     \multicolumn{1}{c}{$n {}^2P_{3/2}$}
				&     \multicolumn{1}{c}{$n {}^2D_{3/2}$} &     \multicolumn{1}{c}{$n {}^2D_{5/2}$}\\
				\hline\\[-0.2cm]
\multicolumn{6}{c}{\bf Li} \\[0.2cm]
				2 & -1.4143&	-1.093&	-1.094 &       &      \\
                & -1.4244&	-1.092&	-1.092 &       &      \\
				  &  -1.43      &  -1.145      &     &        &          \\[0.1cm]
				3 &   -38.086 &46.564&	46.574&	-45.661&	-45.646

      \\
                  &   -38.117 &46.575&	46.585&	-45.641&	-45.645

      \\
			       & -38.62       &  50.80       &       &  -47.75     &        \\[0.1cm]
				4 & -434.846 &	298.507 &	298.485& -224.594& -224.527
      \\
                 & -434.924 &	298.538 &	298.516& -224.510& -224.523
      \\ 
				   & -433.0  & 292.9        &        &  -219.2      &        \\[0.1cm]

\hline\\[-0.2cm]
\multicolumn{6}{c}{\bf Na}\\[0.2cm]

				3 &  -1.388	&-3.118&-3.133&-61.925&-61.947
    \\
				   & -1.389 & -2.985     & -2.998       & -61.24      &  -61.25      \\[0.1cm]
				4 & -27.757&	46.321&	46.154&	14.480&	13.204
     \\
				   &-27.57        &  44.13     & 43.94     &  15.73      &14.46       \\[0.1cm]
				5 & -264.0&-259.5	&-258.4&282.6	&282.5
    \\
			  	 & -264.6 & -254.8      & -253.7      &279.8       &279.7      \\[0.1cm]
\hline\\[-0.2cm]
\multicolumn{6}{c}{\bf K}\\[0.2cm]
                3 &      &        &        &-13.03 & -12.88

     \\
				  &      &       &       & -16.47     &-16.30       \\[0.1cm]
				4 & -2.435	&-5.251&	-5.345&	-160.6&	-159.6
      \\
				  &-2.528       &-5.370         &-5.471        & -171.9      & -170.8       \\[0.1cm]
				5 &  -45.378&  -71.8 & -73.87& 379.7 &379.7
     \\
				   & -45.76       & -76.77       & -78.97       & 357.3      & 358.7     \\[0.1cm]
				6 &  -409.9  & -447.23 &  -456.48& 1389& 1376
      \\
			  	 & -412.5        & -427.1     & -436.2       & 1293      &1282        \\[0.1cm]
				                            
				\end{tabular}
		\end{ruledtabular}
	\end{center}
\end{table}
\begin{table}[ht!]
	\caption{Table~\ref{tab:10} continued.}
    \label{tab:11}
	\begin{center}
		\begin{ruledtabular}
			\begin{tabular}{llllll}
				$n$ &   \multicolumn{1}{c}{$n^2S_{1/2}$}      &   \multicolumn{1}{c}{$n^2P_{1/2}$}   &     \multicolumn{1}{c}{$n^2P_{3/2}$}
				&     \multicolumn{1}{c}{$n^2D_{3/2}$} &     \multicolumn{1}{c}{$n^2D_{5/2}$}\\
				\hline\\[-0.2cm]
\multicolumn{6}{c}{\bf Rb} \\[0.2cm]
                4 &       &        &         & -5.089 &  -4.788
       \\
				 &         &         &        & -6.467      & -6.975         \\[0.1cm]
				5 & -2.645      &  -7.005    & -7.553    &-166.54     & -160.13
       \\
				 & -2.789        & -7.511        & -8.127      &  -188.7      &  -181.4       \\[0.1cm]
				6 & -47.56     &  -142.9    &-156.57   & 57.26    & 51.98

      \\
			     &-48.17      & -151.8        & -166.6        & 71.28       &66.51       \\[0.1cm]
				7 &  -409.7&  -477.10  &  -466.86    &575.7      & 538.2
       \\
				 & -411.7        &-484.5        &-473.8       & 593.8       &554.2        \\[0.1cm]

\hline\\[-0.2cm]
\multicolumn{6}{c}{\bf Cs}\\[0.2cm]
                5 &       &        &       & 3.517 &  4.649
       \\
				  &         &        &         &4.088       &  5.305      \\
                 
				  &         &        &         &5.497        &  5.315      \\[0.1cm]
				6 & -3.299       & -11.98     & -15.24     & -65.54      & -102.4

      \\
      & -3.811       & -12.87     & -16.34     & -65.64      & -102.8

      \\
				  & -3.589       & -14.85       & -17.24        & -70.41       & -99.18      \\[0.1cm]
				7 & -58.07     & -147.4      & -86.19      &-797.7     &-809.0

      \\
                  & -59.26     & -148.5     & -86.44      &-795.8     &-810.9

      \\
				   & -59.67       &-153.4        & -111.7        &-810.4      & -825.8     \\[0.1cm]
				8 & -470.1       & 412.7    &  406.9     & -421.6  &-306.2
      \\
                  & -472.7       & 415.3    &  409.4     & -422.3  &-306.7
      \\
			     & -477.4      & 410.6       & 425.2        &-409.6        &-299.4       \\[0.1cm]

                \hline\\[-0.2cm]

\multicolumn{6}{c}{\bf Fr}\\[0.2cm]
                6 &         &         &        & 2.889       & 6.836
     \\[0.1cm]
				%
				7 &  -2.558      & -10.43   &  -21.51     & -60.98    &-144.2
     \\[0.1cm]
				%
				8 &  -42.95      & -230.6   & -106.5      & -641.0     &-720.3
    \\[0.1cm]
				%
				9 & -319.7    & 198.8       &  360.0  & -192.5    & 51.02
    \\[0.1cm]
				                
			\end{tabular}
		\end{ruledtabular}
	\end{center}
\end{table}

Comparing our results with those in \cite{PhysRevA.23.2397}, there is a good agreement for low-lying $s$-states. For states with larger values of the principal quantum number $n$ and orbital angular momentum $l$, the discrepancy of results becomes more noticeable and is more pronounced for heavier atoms. In the work \cite{PhysRevA.23.2397}, calculations are performed within the Bates-Damgaard method, based on the Coulomb approximation, additionally, an approximate frequency integral calculation is used. We assume that our results are more accurate because our calculation is based on the 'direct' computation of Eq.~(\ref{dStark}). We also presented the calculation results for the fine structure of lithium and the results for the francium atom, which are not available in \cite{PhysRevA.23.2397}.

To assess its impact, we evaluated the effect of the CP correction on the calculated thermal Stark shifts. For the lithium and cesium atoms, every second row presents the values of the shift obtained without this correction. In particular, a basic analysis shows that the influence of the CP correction is minor for light atoms across all states considered. In contrast, for cesium (and heavy atoms in general), this correction is significant, with a more pronounced effect for low-lying states. We also note a significant discrepancy between our results and those of \cite{PhysRevA.23.2397}, which increases with the atomic number $Z$.

The statement regarding the accuracy of our calculations is primarily based on a comparative analysis of the values obtained for static dipole polarizabilities, see previous section. The inaccuracy of our method does not exceed $1\%$ for all the atoms considered. The dc-Stark thermal effect arises from expression (\ref{dStark}) by expanding it in a series in terms of small frequencies $\omega$. Then, in the leading order, an integral of the Planck function with a cubic frequency multiplier arises, whereas the subsequent terms in the expansion (higher powers of frequency) are associated with dynamic temperature corrections to the energy levels. These corrections are significantly less than the leading contribution $\sim (k_{\rm B}T)^4$ due to the greater degree of temperature (recall that $k_{\rm B}T\ll 1$). Corrections of a similar magnitude to the dynamic thermal shifts also arise from other sources, specifically the magnetic-dipole, higher multipole polarizabilities and two-loop diagrams, with these contributions being on the order of $(k_{\rm B} T)^4$ \cite{PhysRevA.108.042801,vzqf-8261}. Thermal shifts themselves are usually calculated purely theoretically, and the comparison is made with experimentally measured atomic polarizabilities \cite{PhysRevA.51.3883,Walls2001,PhysRevA.93.043407,Bai_2020,PhysRevLett.132.233201}. Thus, we conclude that the numerical error of the presented values of Stark shifts does not exceed $1\%$. It should also be noted that, because the core polarizability is nearly frequency-independent, its contribution largely cancels in the dynamic Stark shift of a transition frequency.

\section{Bethe logarithm}
\label{LogB}

The calculation of thermal self-energy is most straightforward for any atomic system. This is primarily due to the absence of ultraviolet divergence, as the Planck distribution function diminishes exponentially with increasing frequency. The zero-temperature loop (the Feynman graph remains consistent with that depicted in Fig.~\ref{fig:2}, yet it presupposes the utilization of a conventional Feynman photon propagator) for many-electron atoms is further complicated by the necessity to sum over the intermediate states of the complete Dirac energy spectrum. In the LDFCP method for the effective one-electron atom, the required set of states is substituted with a quasi-complete, purely discrete spectrum, see section~\ref{meth}. 

Considering the behavior of a single valence electron in a self-consistent field, the one-loop 'zero temperature' contribution can be estimated by calculating the Bethe logarithm for non-relativistic systems \cite{bethe,labzowsky1993relativistic}. In the lowest order for a hydrogen-like atom in the $s$-state, the one-loop self-energy correction can be calculated using the following formula:
\begin{eqnarray}
\label{secorr}
\begin{aligned}
 \Delta E_a^{\mathrm{SE}}=\frac{4Z}{3}\alpha^{3} \left(\frac{19}{30}-2\ln{\alpha{Z}} -\ln{k_0}\right)\left|\Psi(0)\right|^2.
 \end{aligned}
\end{eqnarray}
Here, $\left|\Psi(0)\right|^2$ denotes the square modulus of the wave function evaluated at the nucleus, while $\ln{k_0}$ is referred to as the Bethe logarithm. In the literature it can be found in the form:
 \begin{equation}
   \label{a1}
   \ln{k_{0}}  =  \dfrac{\sum\limits_{n} { |\langle a | {\bf d}| n  \rangle|^2}{(E_{n}-E_{a})}^{3} \ln{\lvert E_{n}-E_{a}\rvert} }{\sum\limits_{n}{ |\langle a | {\bf d}| n  \rangle|^2}{(E_{n}-E_{a})}^{3}}.
\end{equation}
in length guage, and
 \begin{equation}
   \label{a11}
   \ln{k_{0}}  =  \dfrac{\sum\limits_{n} { |\langle a | {\bf p}| n  \rangle|^2}{(E_{n}-E_{a})}^{} \ln{\lvert E_{n}-E_{a}\rvert} }{\sum\limits_{n}{ |\langle a | {\bf p}| n  \rangle|^2}{(E_{n}-E_{a})}^{}}.
\end{equation}
in velocity guage.

Summation by $n$ covers the entire spectrum of an atom. The computation of the Bethe logarithm is complicated by the substantial influence of the continuum states. In the DKB approach, continuous states are integrated with a discrete spectrum, and to achieve a more precise assessment, it is essential to adjust the box size while monitoring the convergence of the results. The combination of LDFCP and DKB methods enables us to estimate the Bethe logarithm.

It is important to note that the quantity in the denominator of Eq.~(\ref{a1}) for a hydrogen atom in the $s$ state is proportional to $\left|\Psi(0)\right|^2$ \cite{bethe,labzowsky1993relativistic}. As a result, accurately reproducing the value of the electron density near the nucleus is crucial. To verify this value, we calculated the hyperfine splitting constant for the state $S_{1/2}$ between two levels of $I\pm 1/2$ using the following formula:
\begin{eqnarray}
 A =-\dfrac{\alpha^{2}}{m_p} \dfrac{ \kappa g_{I}}{ J(J+1)}\int\limits_{0}^{\infty}dr\, g_a(r)\frac{1}{r^2} f_a(r).
\end{eqnarray}
Here, $m_p$ represents the proton mass, $\kappa$ denotes the quantum number of angular momentum, $J$ indicates the total angular momentum of the electron, while $g_I=\mu/(\mu_N I)$ stands for the nuclear $g$-factor, with $\mu_N$ being the nuclear magneton. The values for the nuclear variables were borrowed from \cite{STONE200575}. 

The results of calculations of the hyperfine splitting constant are presented in Table~\ref{tab:8}.
\begin{table}[ht!]
\caption{The hyperfine splitting constant A is given in MHz. The header row lists the atoms for which the calculations were performed, with the corresponding atomic states given in parentheses. The second row lists the values obtained using the LDFCP method, the third row shows the results without the CP correction, and the final row presents the experimental values from \cite{RevModPhys.49.31}. The values in parentheses indicate the relative deviation of the results in percent from the experimental value.}
\resizebox{\columnwidth}{!}{
\label{tab:8}
\begin{tabular}{lllll}  
\hline
  \multicolumn{1}{l}{\textbf{Li} (2s)} & \multicolumn{1}{l}{\textbf{Na} (3s)} & \multicolumn{1}{l}{\textbf{K} (4s)} &
   \multicolumn{1}{l}{\textbf{Rb} (5s)} & \multicolumn{1}{l}{\textbf{Cs} (6s)} 
   
   \\ \hline
   \multicolumn{5}{c}{LDFCP}\\
  $308.6\,\,\, (23)$ & $879.1\,\,\, (0.7)$ & $253.98\,\,\, (10)$ & $3627.12\,\,\, (6)$ & $2407.0\,\,\, (5)$
 \\
  \hline
     \multicolumn{5}{c}{LDF}\\
  $380.01\,\,\, (5.4)$ & $784.93\,\,\, (11)$ & $191.952\,\,\, (17)$ & $2892.73\,\,\, (15)$ & $1966.58\,\,\, (14)$ \\  
     \hline
     \multicolumn{5}{c}{ \cite{RevModPhys.49.31}}\\
  401.7& 885.8& 230.8 & 3417.3& 2298.1\\
  \hline
\end{tabular}
}
\end{table}

As can be seen from the comparison, the result obtained for lithium turned out to be overestimated, the difference from the experimental value is about $23\%$. For other elements, the difference in values is about $10\%$ or lower. In the case of a lithium atom, two core electrons located close to the nucleus should make a significant contribution to the electron density on the nucleus, the value of which is associated with the hyperfine splitting constant. An incorrect assessment of this effect could lead to an underestimated result. In other cases, due to the fact that the number of electrons in an atom increases with increasing Z, this effect no longer makes a strong contribution. So, at work \cite{PhysRevA.100.042506}, using the example of Cs and Rb atoms, it was shown that the contribution of the correction to the core polarization decreases with increasing Z.

In the literature, there are both rigorous calculations (\textit{ab initio}) of the Bethe logarithm for light elements such as hydrogen \cite{PhysRevA.41.1243}, helium \cite{PhysRevA.100.012517}, lithium \cite{PhysRevA.68.042507} and beryllium \cite{PhysRevLett.92.213001} (using explicitly correlated wavefunction models), as well as approximate methods. For example, in \cite{PhysRevA.108.042817} in the mean field approximation, the values of the Bethe logarithm were obtained for elements with $Z=3, 12, 18$.

The results of numerical calculations of the Bethe logarithm for the ground state of some atoms using the LDFCP method combined with the DKB approach are presented in Table~\ref{tab:9}.
\begin{table}[ht!]
\caption{The Bethe logarithm for the ground state of atoms in the alkaline group of elements. The last line contains the values found in the literature with relevant references. The values in each cell are the Bethe logarithms computed using the methods indicated in the column headers: (1) the fully relativistic LDFCP method, (2) the non-relativistic LDF method, (3) the non-relativistic local Kohn-Sham method, and (4) the local core Hartree approach. These results were obtained in two distinct gauges, as defined by Eqs.~(\ref{a1}), (\ref{a11}).}
\label{tab:9}
\begin{ruledtabular}
\begin{tabular}{llllll}  
  \multicolumn{1}{l}{\textbf{Li} (2s)} &\multicolumn{1}{l}{\textbf{Be$^+$} (2s)}& \multicolumn{1}{l}{\textbf{Na} (3s)} & \multicolumn{1}{l}{\textbf{K} (4s)} &
   \multicolumn{1}{l}{\textbf{Rb (5s)}} & \multicolumn{1}{l}{\textbf{Cs (6s)}} 
   \\ \hline
    \multicolumn{5}{c}{LDFCP (relativistic)}\\
  $5.23$&$6.09$& $6.26$  & $7.41$& $8.34$ & $8.9$ \\
 \hline
   \multicolumn{5}{c}{LDF (non-rel)}\\
    \multicolumn{5}{c}{lenght guage}\\
  $5.1922$& $5.710$ &$7.6332$ &$8.6595$  & $9.9185$  & $10.650$  \\
   \multicolumn{5}{c}{velocity guage}\\
     $5.1921$& $5.702$ &$7.6332$ &$8.6595$  & $9.9185$  & $10.650$  \\
   \hline
   \multicolumn{5}{c}{local Kohn-Sham (non-rel)}\\
    \multicolumn{5}{c}{lenght guage}\\
  $5.1312$&$5.673$&$7.6269$ & $8.6596$ & $9.9197$  & $10.651$  \\
      \multicolumn{5}{c}{velocity guage}\\
  $5.1307$&$5.669$&$7.6269$ & $8.6595$ & $9.9197$  & $10.651$  \\    
     \hline
   \multicolumn{5}{c}{local CH (non-rel)}\\
    \multicolumn{5}{c}{lenght guage}\\
  $5.1871$&$5.710$&$7.6453$ & $8.672$  & $9.9263$   & $10.656$  \\
  \multicolumn{5}{c}{velocity guage}\\
   $5.1871$&$5.710$&$7.6452$ & $8.672$  & $9.9263$   & $10.656$  \\
     \hline
  5.17817 \cite{PhysRevA.68.042507} &5.75167\cite{PhysRevLett.92.213001}& 7.770 \cite{PhysRevA.108.042817}&&&\\
\end{tabular}
\end{ruledtabular}
\end{table}

Due to a substantial discrepancy between our LDFCP value for the sodium Bethe logarithm and that in \cite{PhysRevA.108.042817}, we performed calculations using several alternative methods. Specifically, we employed the local Dirac-Fock, local Kohn-Sham, and local core-Hartree (CH) approaches in the non-relativistic limit, neglecting the core polarization correction. The latter serves to unambiguously verify the correctness of the calculations. Furthermore, the calculations were carried out in two different gauges, length and velocity, according to Eqs.~(\ref{a1}), (\ref{a11}).

According to this calculation, the values obtained are perfectly consistent with those available in the literature \cite{PhysRevA.68.042507,PhysRevLett.92.213001,PhysRevA.108.042817}. Thus, we conclude that the semi-empirical Core Polarization correction significantly spoils the calculation of the Bethe logarithm. This is due to the fact that the semi-empirical potential $V_{\rm CP}$, see Eq.~(\ref{cp}), leads to incorrect behavior of wave functions on the nucleus (at $r\rightarrow 0$) and therefore incorrect calculation of the quantity in the denominator of Eq.~(\ref{a1}). A likely major source of error in the LDFCP calculations is also the neglect of core polarizability, as there is no precisely defined value for this quantity in the literature. This suggests that this contribution should also be considered for an accurate evaluation of the Bethe logarithm. The Bethe logarithm calculations for Rb and Cs serve mainly to test the method on complex many-electron systems; they should not be considered reliable \textit{a priori} for evaluating the self-energy correction in these cases. The data in the table clearly demonstrate that the relativistic LDFCP method is inadequate for treating complex QED effects related to the electron density at the nucleus.

%
%
\section{Conclusion and discussion}

This paper presents the LDFCP method, which allows us to obtain the energy spectrum and wave functions of the effective one-particle Hamiltonian. Based on this, calculations of atomic characteristics were carried out for monovalent atoms of the alkali group. In particular, the values of atomic polarizabilities and the corresponding thermal shifts of $nS$, $nP_{j}$, $nD_{j}$ atomic energy levels, static polarizability values for highly excited states of Rb and Cs atoms, as well as the value of the Bethe logarithm for the ground state of alkali atoms were obtained. The numerical values of the quantities calculated by the LDFCP method are in good agreement with the data from the literature.

The calculation of atomic polarizabilities via the LDFCP method directly enables the determination of thermal Stark shifts for different states of alkali atoms. Although our numerical results show satisfactory agreement with those of \cite{PhysRevA.23.2397}, substantial differences remain. The main reasons for this are twofold: (a) a more accurate treatment of the atomic level fine structure in our approach, and (b) the use of a 'direct' frequency integration technique, in contrast to the approximate method used in \cite{PhysRevA.23.2397}. Given their fundamental relevance to atomic clocks, we consider our numerical values for the thermal Stark shifts to be more precise (a detailed discussion is provided in the main text).

The greatest advantage of the method is the possibility of obtaining a complete Dirac spectrum of the atomic electron (the actual spectrum is replaced by a purely discrete pseudo-spectrum). This circumstance allows us to carry out the necessary summation over intermediate states, which arises within the framework of the perturbation theory of second and higher orders. We evaluate the accuracy of the method by comparing the Dirac energies and dipole matrix elements for monovalent atoms. In calculating the electric dipole polarization, we found that the method is accurate to better than $1\%$. Even for highly excited states the accuracy of the calculations is quite high. Another advantage is the comparatively low computational and time costs of the LDFCP method.

The most challenging for computations is logarithm Bethe. In calculating this quantity, it was found that a reasonable agreement existed for the ground state of the lithium atom, whereas for the other monovalent elements the discrepancy between the obtained values and the accepted ones is considerable. Therefore, given the trend of increasing deviation for heavier atoms, the LDFCP values cannot be considered reliable in relativistic calculations. To verify this, the core polarization correction was excluded, and the calculation was switched to the non-relativistic framework of the LDF method. The semi-empirical CP correction was omitted because it causes unphysical behavior of the wave function at the nucleus, resulting in an erroneous denominator in expression (\ref{a1}). Upon reverting to the LDF method, values for the Bethe logarithm were obtained that are in good agreement with data available in the contemporary literature. A direct comparison of the LDF and LDFCP results justifies the use of the CP correction only for light atomic systems. To further verify this, numerical calculations of the Bethe logarithm were performed using the local Kohn-Sham and local Core-Hartree approaches in both the length and velocity gauges. These calculations showed excellent agreement with the LDF method.

The primary conclusion of this work is that the LDFCP method is applicable for calculating properties that do not depend on the wave function's behavior at the nucleus. For such properties, the method yields results in good agreement with known values, allowing for reliable, high-accuracy predictions. The unphysical behavior of the CP correction at the origin merely ensures the correct order of magnitude, thereby defining the limits of the LDFCP method's applicability.

\acknowledgments
The work of the authors was supported by grants from the Foundation for the Advancement of Theoretical Physics and Mathematics "BASIS"\, Nos.~23-1-3-31-1 and 25-1-2-18-1.
%
%

%
\bibliographystyle{apsrev4-1}
\bibliography{ref}
%
%
%

\end{document}